\documentclass[12pt]{article}
\usepackage{graphicx}
\usepackage{multirow}

\newcommand\pubnumber{CIPANP2015-DeAlmeidaDias}
\newcommand\pubdate{\today}

\def\edin{School of Physics and Astronomy, University of Edinburgh, Edinburgh, United Kingdom}

\def\Title#1{\begin{center} {\Large #1 } \end{center}}
\def\Author#1{\begin{center}{ \sc #1} \end{center}}
\def\Address#1{\begin{center}{ \it #1} \end{center}}

\newcommand\pubblock{\rightline{\begin{tabular}{l} \pubnumber\\
         \pubdate  \end{tabular}}}
\newenvironment{Abstract}{\begin{quotation}  }{\end{quotation}}
\newenvironment{Presented}{\begin{quotation} \begin{center} 
             PRESENTED AT\end{center}\bigskip 
      \begin{center}\begin{large}}{\end{large}\end{center} \end{quotation}}

\def\beq{\begin{equation}}
\def\eeq#1{\label{#1}\end{equation}}
\def\eeqn{\end{equation}}

\def\beqa{\begin{eqnarray}}
\def\eeqa#1{\label{#1}\end{eqnarray}}
\def\eeqan{\end{eqnarray}}

\let\bar=\overbar

\def\Dslash{\not{\hbox{\kern-4pt $D$}}}
\def\dslash{\not{\hbox{\kern-2pt $\del$}}}

\def\msb{{\bar{\ssstyle M \kern -1pt S}}}

\begin{document}
\begin{titlepage}
\pubblock

\vfill
\Title{Prospects for Higgs properties measurements at future colliders}
\vfill
\Author{ Flavia De Almeida Dias\\ On behalf of the ATLAS and CMS Collaborations}
\Address{\edin}
\vfill
\begin{Abstract}
The LHC Run-1 was very successful and included the discovery of a new
particle with mass of about 125 GeV compatible with the 
Higgs boson predicted by the Standard Model. The prospects for Higgs physics 
at the high-luminosity LHC and at future colliders are presented. In particular, the 
ultimate precision attainable for the couplings measurements of the 125 GeV particle 
with elementary fermions and bosons is discussed along with prospects for self-coupling 
measurements, for the ATLAS and CMS detectors at the upgraded LHC.\end{Abstract}
\vfill
\begin{Presented}
Twelfth Conference on the Intersections of Particle and Nuclear Physics -- 
CIPANP2015\\
Vail, USA,  May 19 - 24, 2015


\end{Presented}
\vfill
\end{titlepage}
\def\thefootnote{\fnsymbol{footnote}}
\setcounter{footnote}{0}

\section{Introduction}

The first operational period of the Large Hadron Collider (LHC), from 2010 - 2012, was very
successful, including the discovery of a new particle with mass of about 125 GeV
compatible, within uncertainties, with the Higgs boson predicted by the Standard Model (SM). 
Precise measurements of the properties of this boson, and the discovery of new physics
beyond the Standard Model, are primary goals of the future LHC programme. 

The High-Luminosity LHC (HL-LHC) is the planned upgrade of the LHC, and has an aim to enable 
precise measurements of the Higgs boson properties. 
In the SM, all properties of the Higgs boson are defined once its mass is known. However, this model has
many open questions such as the hierarchy problem and the nature of dark matter~\cite{BSM,DM}. 
Many alternative theories
addressing these issues predict that the SM Higgs couplings are modified, or the existence of
additional Higgs bosons~\cite{MultiHiggs}. 
%
The present LHC programme is expected to deliver a total integrated luminosity of about 300 fb$^{-1}$ by
the year 2022. The peak instantaneous luminosity will be in the range from 2 to 
$3 \; \times \; 10^{34}$ cm$^{-2}$s$^{-1}$. 
This data is assumed to have an
average number of pile-up events per bunch crossing, which is denoted here by $\mu_{PU}$, of 50 - 60. 
The $\textrm{HL-LHC}$ is scheduled to start in 2026, and
will deliver a total luminosity of about 3000 fb$^{-1}$, at a peak luminosity of 
$5 \; \times \; 10^{34} \; \textrm{cm}^{-2} \textrm{s}^{-1}$,
with a value of $\mu_{PU}$ = 140.

These LHC accelerator upgrades require major upgrades of the ATLAS~\cite{ATLAS_JINST} and 
CMS~\cite{CMS_JINST} detectors in order to maintain 
the current performance capabilities under new challenging operating conditions. The upgrade of those detectors 
is absolutely central to the exploitation of the rich physics potential provided by the HL-LHC datasets. 
The purpose of this document is to summarise some highlights of
the future physics potential of the ATLAS and CMS experiments at the HL-LHC, in the
context of measurements of the SM Higgs boson.
%
%

The methodology used by the ATLAS and CMS collaborations to obtain the expected sensitivity 
at the HL-LHC is different. 
On ATLAS, the analyses use particle-level simulation. The response of the upgraded ATLAS detector to the final
state particles is parameterised and applied to the output of the Monte Carlo generators. The parameterised
response of the detectors expected at HL LHC conditions is obtained from studies of
ATLAS full simulation events samples~\cite{ATL-PHYS-PUB-2013-009, ATL-PHYS-PUB-2013-004}. 
All requirements on the final state objects are applied on
the parameterised properties.
At high luminosity, the theory uncertainties start playing an important role in the total uncertainty. 
For that reason, the ATLAS results are quoted for two scenarios: one in which all systematic uncertainties 
are considered, including the theory uncertainties, and another where the theory uncertainties are not included.

On CMS, the projections are based on the assumption that the planned upgrades of the CMS detector will
achieve the goal of mitigating the increased radiation damage and complications arising from
higher luminosity and higher pile-up. With this primary assumption, existing public results
based on current data are extrapolated to higher energy and luminosities. In most cases, the
analyses are assumed to be unchanged~\cite{CMS_HL}. 
The results are quoted in two different scenarios: one in which the Run-1 analysis systematics are used, and 
another in which the systematic uncertainties are reduced by 50\%, under the assumption that the most relevant 
uncertainties, which are related to theory predictions, will be reduced in the future, and that with more 
data and improved analysis techniques, the sources of uncertainties will be better understood.

\section{Selected results on individual channels}

This section will highlight three channels, which were not observed at the LHC Run-1
at 5$\sigma$ significance due to their overwhelming backgrounds or low branching ratio, but are 
of interest in Run-2 and the HL-LHC: $\textrm{H} \rightarrow b\bar{b}$, 
H $\rightarrow$ Z$\gamma$ and H $\rightarrow$ $\mu\mu$.
The results are quoted as relative uncertainties in the signal strength ${\mu} = {\sigma_{obs}}/{\sigma_{SM}}$,
where $\sigma_{obs}$ is the observed cross section and $\sigma_{SM}$ is the SM cross section for a Higgs boson of
$m_H = 125$ GeV. As all the results assume a SM Higgs, all the signal strengths are presumed to be ${\mu} =1$.

At the LHC, the largest Higgs boson production cross section is through gluon-gluon fusion ($ggF$).
%
Other production mechanisms include, in descendant cross section, vector boson fusion (VBF), 
associate production with a 
vector boson (WH, ZH), and associate production with a pair of 
top quarks ($t\bar{t}$H).

The SM Higgs boson decays to a pair of bottom quarks 
with the largest branching ratio (58\%). 
However, the search for H $\rightarrow b\bar{b}$ in the $ggF$ production mode suffers from a huge background from
direct $b$-quark production. For this reason, the experiments focus on searches in the associate 
production modes (WH, ZH and $t\bar{t}$H).
The ATLAS analysis looked at Higgs bosons in the WH and ZH productions, 
where the Z and W boson decay leptonically 
($\textrm{ZH}\rightarrow e^+ e^-b\bar{b},\; \mu^+\mu^-b\bar{b}$ 
and $\textrm{W}^{\pm}\textrm{H} \;\rightarrow e^{\pm}\nu_e b\bar{b}, \; \mu^{\pm}\nu_{\mu}b\bar{b}$)~\cite{ATLAS_bb}. The channel 
$\textrm{Z}\rightarrow \nu \nu b\bar{b}$ was not considered due to the lack of studies on the feasibility 
of a low threshold missing transverse momentum ($E_{\mathrm{T}}^{\mathrm{miss}}$) trigger in the high pile-up environment. 
The CMS experiment results for the H $\rightarrow$ $b\bar{b}$ include the associated productions ZH, WH and 
$t\bar{t}$H, using all the channels considered in the Run-1 analysis~\cite{CMS_VHbb, CMS_ttHbb}. 
The expected uncertainties on ${\mu}$ are shown in Table~\ref{tab:channels}.
%
%
The CMS results show better sensitivity due to the inclusion of more channels in the HL-LHC sensitivity studies, 
which was possible owing to the methodology of extrapolating Run-1 results to higher energy and luminosities.
It also does not consider pile-up degradation, which would affect specially the channels relying on the $E_{\mathrm{T}}^{\mathrm{miss}}$.

The H $\rightarrow \textrm{Z} \gamma$ decay is interesting because in the SM the decay proceeds entirely
via loops, predominantly involving heavy charged particles, a property that makes it sensitive to possible
new heavy states. The measurement of the H $\rightarrow \textrm{Z} \gamma$ decay rate can also provide 
insight into physics beyond the SM. 
The projection of  H $\rightarrow \textrm{Z} \gamma$ considered the $ggF$, VBF, VH and $t\bar{t}$H production modes. 
The main background is  Z$\gamma$ with $\gamma$ being a radiative photon. Due to the very low branching ratio of this decay in the SM, the statistical uncertainty in the signal strength $\mu$ is more important than the systematic uncertainties.
The expected uncertainty on ${\mu}$ is shown in Table~\ref{tab:channels}~\cite{ATLAS_Zg,CMS_HL}. At the HL-LHC,
the signal strength for this decay can be measured with uncertainties around 20-30\%.

The Higgs boson decay to a pair of muons is rare in the SM. It allows the coupling to second generation fermions to be 
probed, and can contribute to mass measurements due to the high resolution of the reconstructed $\mu \mu$
invariant mass.
This decay channel also has great interest for beyond the SM, as many theories predict Higgs 
decaying to muons more frequently than the SM expectation. 
The production modes considered in the H$\rightarrow \mu \mu$ results are $ggF$, VBF, VH and $t\bar{t}$H.
The main backgrounds are
Z+jets and $t\bar{t}$ events. The expected uncertainty on ${\mu}$ is shown in Table~\ref{tab:channels}~\cite{ATLAS_ECFA,CMS_HL}. At the HL-LHC, 
the signal strength for this decay can be measured with uncertainties around 20\%.

\begin{table}[h!]
\begin{center}
\begin{tabular}{cc|cc|cc}  
& & 
\multicolumn{2}{c}{ATLAS} &
\multicolumn{2}{|c}{CMS} \\
\hline
Channel & $\mathcal{L}$ (fb$^{-1}$) 
& w/ theory sys & w/o theory sys & Run-1 sys & reduced sys  \\
\hline
\multirow{2}{*}{ H $\rightarrow$ $b\bar{b}$} & 300 & $\pm$ 0.26 & $\pm$ 0.25 & $\pm$ 0.14  & $\pm$ 0.11 \\
 & 3000 & $\pm$ 0.13 & $\pm$ 0.11 & $\pm$ 0.07 & $\pm$ 0.05 \\
\hline
\multirow{2}{*}{ H $\rightarrow$ Z$\gamma$} & 300 & $\pm$ 0.46 & $\pm$ 0.44 & $\pm$ 0.62  & $\pm$ 0.62 \\
 & 3000 & $\pm$ 0.30 & $\pm$ 0.27 & $\pm$ 0.24 & $\pm$ 0.20 \\
\hline
\multirow{2}{*}{ H $\rightarrow$ $\mu\mu$} & 300 & $\pm$ 0.39 & $\pm$ 0.38 & $\pm$ 0.42  & $\pm$ 0.40 \\
 & 3000 & $\pm$ 0.16 & $\pm$ 0.12 & $\pm$ 0.20 & $\pm$ 0.14 \\
\hline
\end{tabular}
\caption{Precision of the signal strength measurement per decay mode for a SM-like
Higgs boson. These values are obtained at 
$\sqrt{s} = 14$ TeV using an integrated dataset of 300 and 3000 fb$^{-1}$. 
Reported values are the uncertainties on the measurements (assuming signal strength 
${\mu} = 1$) estimated under
different systematic uncertainty scenarios, as described earlier in the text. 
}
\label{tab:channels}
\end{center}
\end{table}

\section{Higgs coupling fit}

To measure the signal strength in units of the SM expectation, a coupling fit is performed.
The measurements of coupling scale factors are implemented using a leading-order tree-level motivated
framework~\cite{coupling}.  The $\kappa$-framework is based on the assumptions that there is a single resonance
of mass $m_H=125$ GeV, with a narrow width, 
and only modifications of coupling strengths are considered, while the tensor structure of the Lagrangian
is assumed to be the same as that in the SM. This assumes in particular that the observed
state is a CP-even scalar.
The Higgs boson coupling scale factors are determined from a combined fit to all the channels available, 
where the product $\sigma \; . \; \textrm{BR}$ of cross section and branching ratio for all contributing
Higgs signal channels is expressed as function of the coupling scale factors $\kappa_i$. The parameterisations
used follow the recommendations in~\cite{coupling}.
%
Table~\ref{tab:kappa} shows the uncertainties obtained on $\kappa_i$ for an integrated dataset of 300 fb$^{-1}$
and 3000 fb$^{-1}$. The expected precision ranges from 5 - 23\% for 300 fb$^{-1}$
and 2 - 12\% for 3000 fb$^{-1}$~\cite{CMS_HL, ATLAS_coupling}. 
The measurements are limited by systematic uncertainties in the cross section,
which are included in the fit for the signal strength. The statistical uncertainties on $\kappa_i$ are below
one percent. 

\begin{figure}[!h]
\centering
\includegraphics[width=0.41\textwidth]{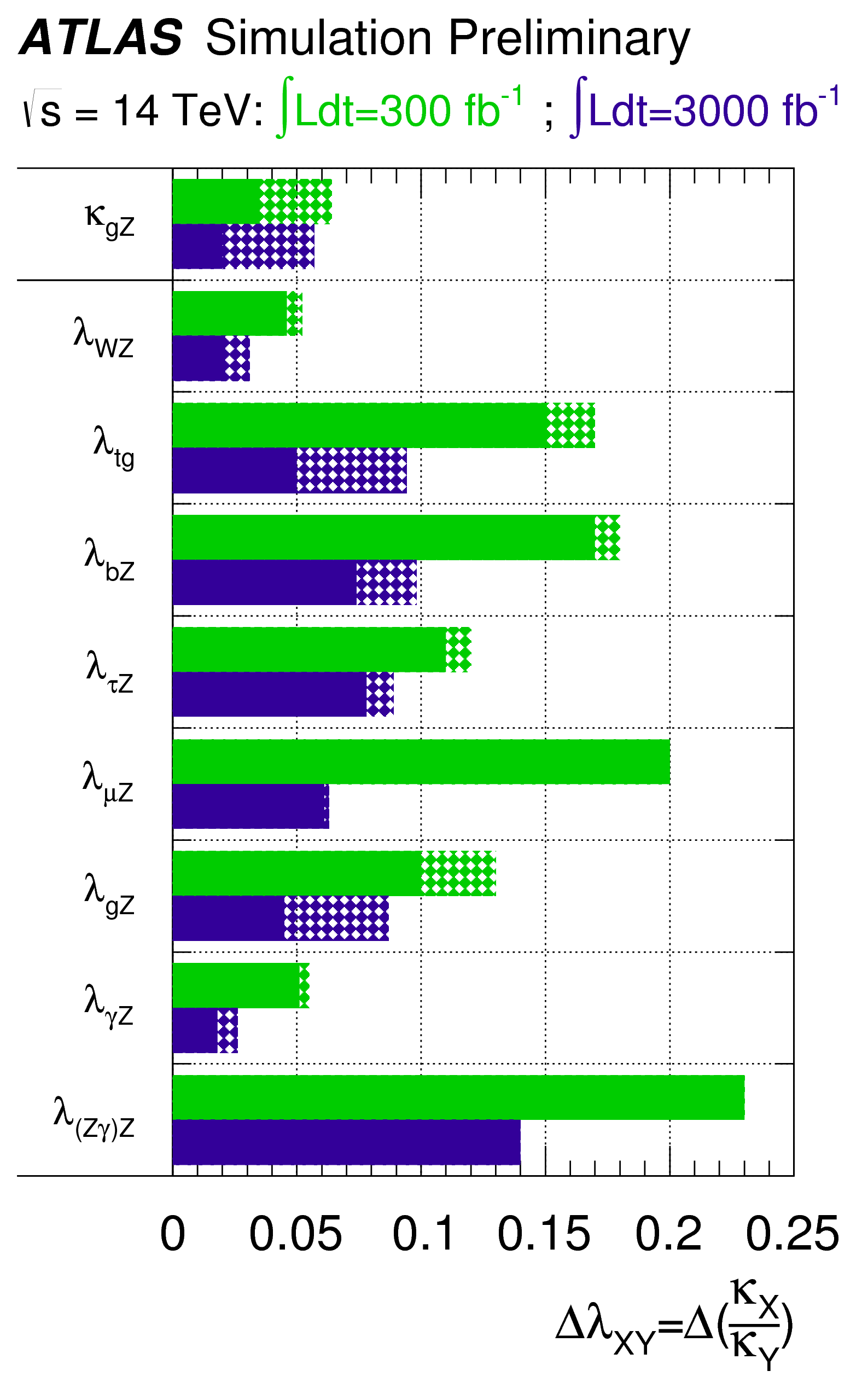}
\includegraphics[width=0.48\textwidth]{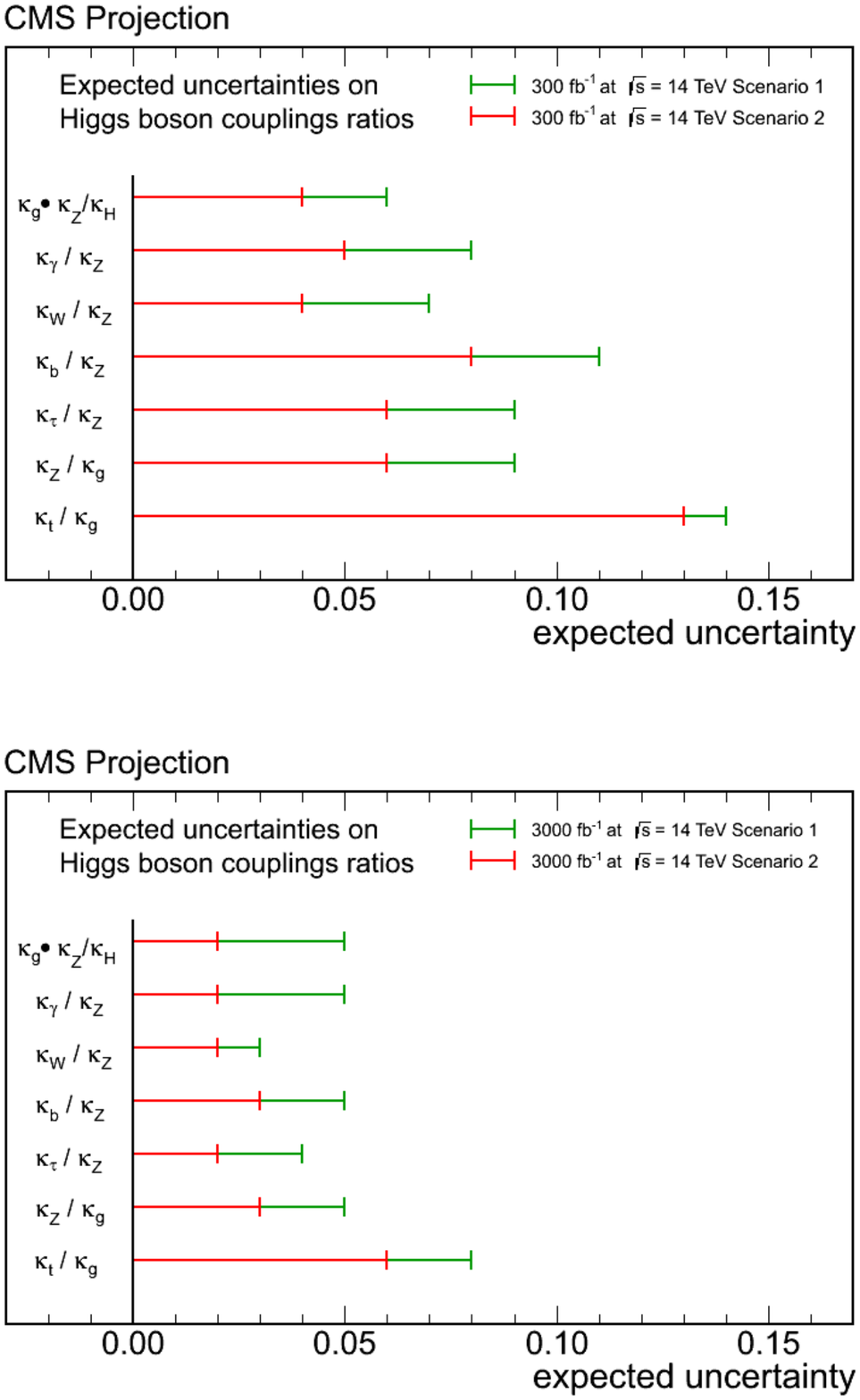}
\caption{
Relative uncertainty expected for the determination of coupling scale factor ratios 
$\lambda_{XY} = \kappa_{X}/\kappa_{Y}$
in a generic fit without assumptions, assuming a SM Higgs boson with a mass of 125 GeV.
Left: ATLAS predictions with 300 fb$^{-1}$ or 3000 fb$^{-1}$ of 14 TeV LHC data. The hashed areas indicate the increase of the estimated error due to current theory systematic uncertainties~\cite{ATLAS_coupling}.
Right: CMS predictions, for 300 fb$^{-1}$ (top) and 3000 fb$^{-1}$ (bottom), with reduced and Run-1 systematics 
scenarios as described in the introduction~\cite{CMS_HL}. 
}
\label{fig:coup}
\end{figure}

A more model-independent way to perform the coupling fit is to remove the assumption on 
the total width of the Higgs boson resonance. In this method, only the ratio of
coupling scale factors can be determined at the LHC. In this case,  
$\sigma \; . \; \textrm{BR}$  for all signal channels
are a function of products of ratios 
$\lambda_{XY} = \kappa_X / \kappa_Y$ of coupling scale factors giving the proportionality
 $\sigma \;. \; \textrm{BR} \sim \lambda^2_{iY}\; .\; \kappa^2_{YY'} \;.\; \lambda^2_{fY} $, 
 where $\kappa_{YY'}=\kappa_Y \; . \; \kappa_{Y'} \; / \kappa_H$ is a suitable chosen overall scale
parameter common to all signal channels and $\lambda_{iY}$ and $\lambda_{fY}$ are the coupling scale factor ratios involving
the initial and final state particles, respectively. In addition to avoid the assumption on the total width,
ratios of coupling scale factors also have the advantage that many experimental and theoretical systematic
uncertainties cancel (such as the uncertainty in the integrated luminosity).
The expected precision on the ratios of coupling parameters are given in Table~\ref{tab:lambda} and Figure~\ref{fig:coup}.
The parameter $\lambda_{\gamma Z}$ can be used as a probe of new charged particles contributing to 
H$\rightarrow \gamma \gamma$ decay loop when compared to the H$\rightarrow$ ZZ$^{*}$, and 
$\lambda_{tg}$ as a probe of new coloured particles contributing to $ggF$ production loop when compared to 
$t\bar{t}$H/$t$H production.

\begin{table}[h!]
\begin{center}

\begin{tabular}{@{}c|c|ccccccc@{}}  
& $\mathcal{L}$ (fb$^{-1}$) & $\kappa_{\gamma}$ & $\kappa_{W}$ & $\kappa_{Z}$ & $\kappa_{g}$ & $\kappa_{b}$ & $\kappa_{t}$ & 
$\kappa_{\tau}$ \\ 
\hline
ATLAS & 300 & [9,9]	& [9,9] &	[8,8]	& [11,14]	& [22,23] &	[20,22]	& [13,14]	\\
CMS & 300 & [5,7]& [4,6] &[4,6] & [6,8]	&[10,13] &	[14,15]&[6,8] \\ 
\hline
ATLAS & 3000 & [4,5] & [4,5] & [4,4] & [5,9] & [10,12] & [8,11] & [9,10] \\
CMS & 3000 & [2,5] & [2,5] & [2,4] & [3,5] & [4,7] & [7,10] & [2,5] \\ 
\hline
\end{tabular}

\caption{Expected precision on Higgs boson coupling scale factors with 300 and 3000 fb$^{-1}$ of
$\sqrt{s}=14$ TeV data. Numbers in brackets are the relative uncertainties (in \%) on couplings for two different scenarios 
of systematic uncertainties: ATLAS [w/o theory, w/ theory], CMS [reduced, Run-1],  as described in the introduction.
}
\label{tab:kappa}
\end{center}
\end{table}

\begin{table}[h!]
\begin{center}

\begin{tabular}{@{}c|c|ccccccc@{}}  
& $\mathcal{L}$ (fb$^{-1}$) & $\kappa_{gZ}$ & $\lambda_{\gamma Z}$ & $\lambda_{WZ}$ & $\lambda_{bZ}$ & $\lambda_{\tau Z}$ 
& $\lambda_{gZ}$ & $\lambda_{tg}$ \\ 
\hline
ATLAS & 300 & [2,6] & [2,3] & [2,2] & [7,10] & [8,9] & [5,9] & [5,9] \\
CMS & 300 & [4,6] & [5,8] & [4,7] & [8,11] & [6,9] & [6,9] & [13,14] \\
\hline
ATLAS & 3000 & [2,6] & [2,3] & [2,2] & [7,10] & [8,9] & [5,9] & [5,9] \\
CMS & 3000 &[2,5] & [2,5] & [2,3] & [3,5] & [2,4] & [3,5] & [6,8] \\
\hline
\end{tabular}

\caption{Expected precision on the ratios of Higgs couplings 
with 300 and 3000 fb$^{-1}$ of $\sqrt{s}=14$ TeV data. 
Numbers in brackets are the relative uncertainties (in \%) on couplings for two different scenarios 
of systematic uncertainties: ATLAS [w/o theory, w/ theory], CMS [reduced, Run-1],  as described in the introduction.
}

\label{tab:lambda}
\end{center}
\end{table}

\section{HH production}

One of the exciting prospects of the HL-LHC is the observation of pairs of Higgs boson (HH), with cross
section of 40.7 fb at NNLO~\cite{HHNNLO} at $\sqrt{s} = 14 $ TeV. 
The final states studied are HH $\rightarrow b\bar{b} \gamma \gamma$ and HH $\rightarrow b\bar{b}$WW: 
the first would yield only 320 expected
events at 3000 fb$^{-1}$, with relatively clean signature; the second about 30,000 events at $\textrm{3000 fb}^{-1}$, but with large background
yields. Other final states under consideration are HH $\rightarrow b\bar{b}b\bar{b}$ and 
HH $\rightarrow b\bar{b}\tau \tau$~\cite{ATLAS_HH_bbgg}.

\begin{figure}[!h]
\centering
\includegraphics[width=0.52\textwidth]{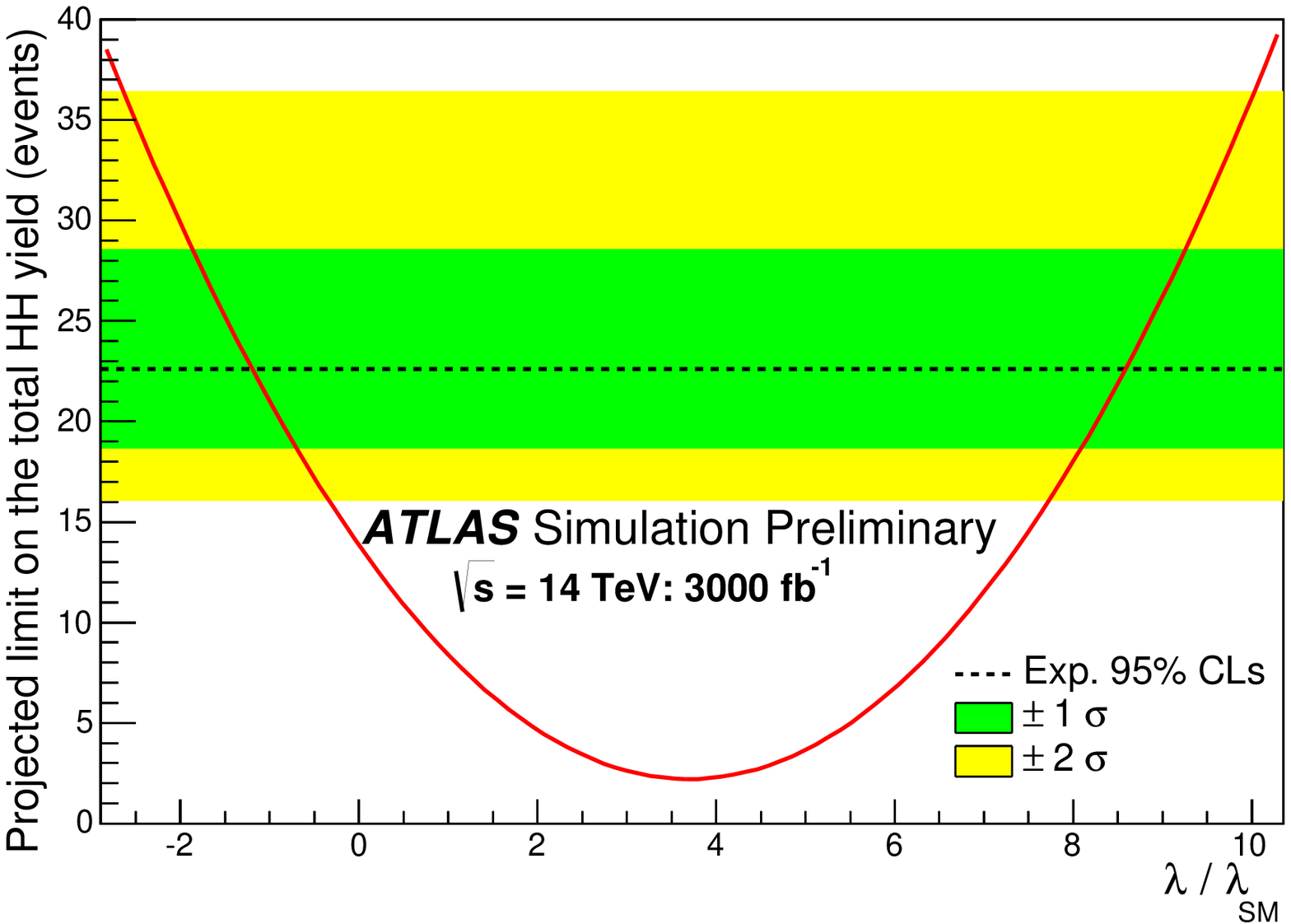}
\includegraphics[width=0.47\textwidth]{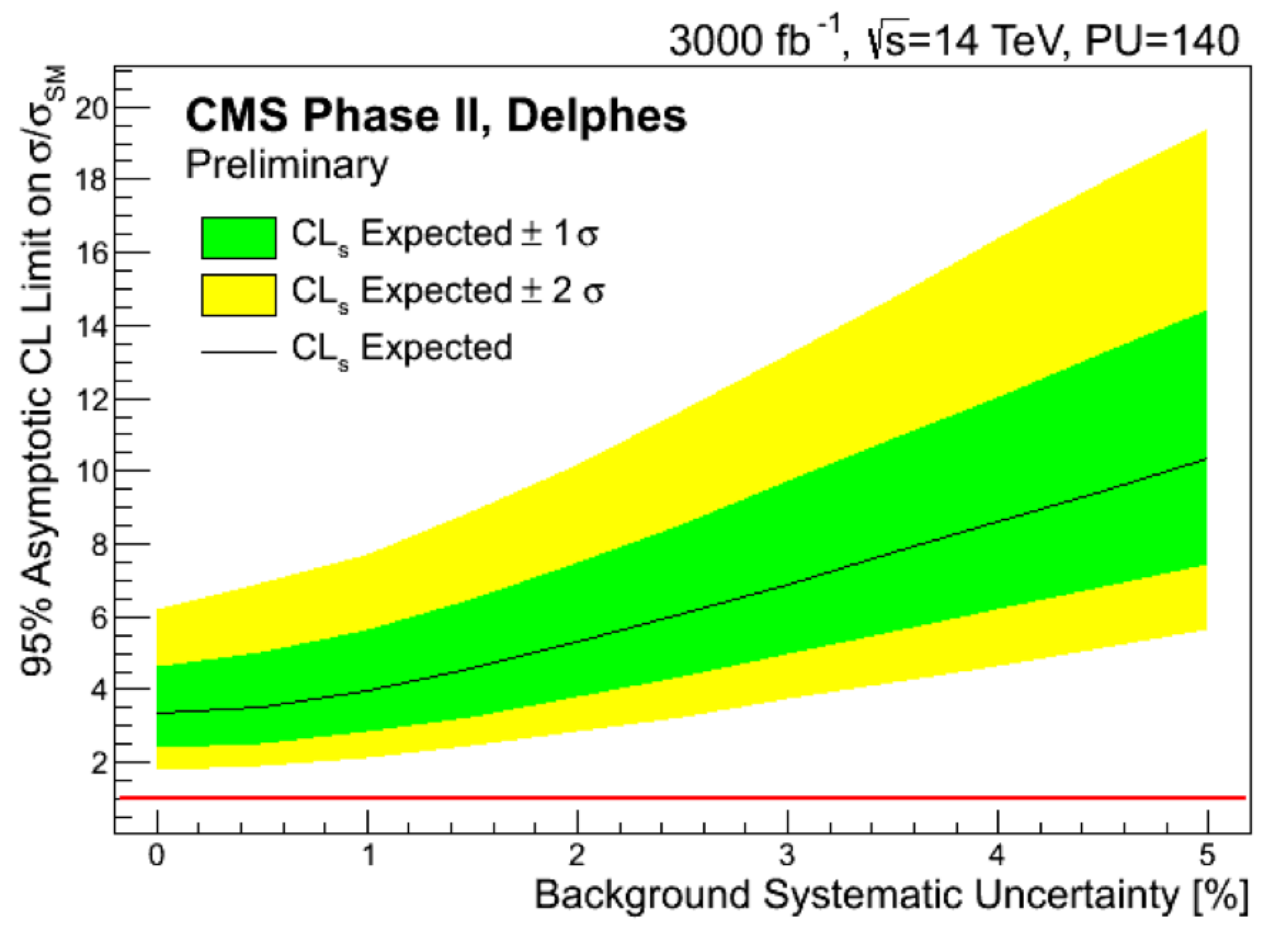}
\caption{Left: ATLAS expected limit (dashed line) on the size of an additional HH contribution summed with the expected SM HH yield, overlaid on the number of predicted total (box + self-coupling) HH events as a function of the $\lambda/\lambda_{SM}$ ratio (solid line)~\cite{ATLAS_HH_bbgg}. Right: The CMS expected limit, for 3000 fb$^{-1}$ HL-LHC, as a function of the background systematic uncertainty, for the H $\rightarrow bb$WW search~\cite{CMS_bbww}. 
}
\label{fig:HH}
\end{figure}

Figure~\ref{fig:HH} (left) shows, for the HH $\rightarrow b\bar{b} \gamma \gamma$ analysis, the ATLAS 
expected limit on the size of an additional HH contribution to the expected SM
HH yield, overlaid on the number of predicted total (box + self-coupling) HH events as a function 
of the $\lambda/\lambda_{SM}$ ratio. Values of $\lambda/\lambda_{SM} < -1.3$ and $\lambda/\lambda_{SM} > 8.7$ can be excluded.
For this channel, the CMS experiment made performance scans on $\sigma$(HH) 
measured uncertainty in different scenarios. 
Figure~\ref{fig:HH} (right) shows the CMS HH $\rightarrow b\bar{b}$WW analysis expected limits, as a function of the background systematic uncertainty.

\section{Summary}

 Detailed characterisation of the Higgs boson is essential to elucidate the nature of
 the electroweak symmetry breaking mechanism, and it may give insights as to where the next discoveries may occur. 
One of the most important components of this quest is the exploitation of the full potential of the
upgraded LHC. The HL-LHC is going to be a Higgs factory, producing a total of 170 million Higgs bosons and 
121 thousand HH events with accumulated dataset of 3000 fb$^{-1}$ at $\sqrt{s} = 14$ TeV. This is an unique opportunity
to study the SM Higgs rare decays and couplings, spin and parity properties, Higgs pair production, and to investigate
if there are any hints of beyond SM in the Higgs sector, such as more Higgs-boson-like particles (predicted in MSSM, 2HDM models), 
if the Higgs boson couples to Dark Matter, or is composite. 

Several new Higgs boson production and decay modes can be observed by the ATLAS and CMS detectors with
3000 fb$^{-1}$ at the HL-LHC compared to a sample of 300 fb$^{-1}$
that will be accumulated before the Phase-II upgrades, and the precision of all channels can be improved. 
The projected measurements of cross section times branching ratio using the combination of
all channels studied are used to derive the expected precision on Higgs boson couplings to fermions
and bosons, and on the ratios of couplings. 


\end{document}